\documentstyle[sprocl]{article}

\def\Journal#1#2#3#4{{#1} {\bf #2}, #3 (#4)}

\def\NPB{{\em Nucl. Phys.} B}
\def\PLB{{\em Phys. Lett.}  B}
\def\PRL{\em Phys. Rev. Lett.}
\def\PRD{{\em Phys. Rev.} D}

\def\APJL{\em Astroph. J. Lett.}
\def\JL{\em JETP Lett.}
\def\Nat{\em Nature}
\def\MN{\em MNRAS}
\def\AA{\em Astron. Astroph.}
\def\vph{\varphi}

\def\be{\begin{equation}}
\def\ee{\end{equation}}
\def\bea{\begin{eqnarray}}
\def\eea{\end{eqnarray}}

\begin{document}

\title{INFLATION FOR LARGE SCALE STRUCTURE}

\author{ A.A. STAROBINSKY}

\address{Landau Institute for Theoretical Physics, Kosygina 2, Moscow, 117940,
Russia}


\maketitle\abstracts{
Two extensions of ideas lying in the basis of the inflationary scenario
of the early Universe and their effect on the large scale structure of the
present-day Universe are discussed. The first of them is 
the possibility of fast phase transitions in physical fields other than an 
inflaton scalar field during inflation and nor far from the end of it.
This results in the appearance of specific features in the inflaton effective
potential which, in turn, lead to the generation of localized spikes in the 
primordial perturbation spectrum. At present, there seems to exist one 
scale in the spectrum, $k=0.05h$ Mpc$^{-1}$, around which we might see 
something of this type.
The second one is the possibility that we are living at the beginning of a
new inflation-like era now. Then observations of clustering of high-redshift
objects can provide information sufficient for the unambiguous
determination of the effective potential of a corresponding present inflaton
scalar field.}
  
\section{Introduction}

The main importance of the inflationary scenario of the early Universe
for the theory of large scale structure in the present-day Universe is that
the former scenario predicts (in its simplest realizations) an approximately
flat, or scale-invariant, spectrum ($n(k)\equiv {d\ln P_0(k)/
d\ln k}\approx 1$) of initial adiabatic perturbations.
By the simplest realizations I mean, as usually, inflationary models with
one effective slow-rolling scalar (inflaton) field. Of course, the physical
nature of the inflaton may be completely different in these models, but it 
does not matter for observations, in particular, for the large scale 
structure. This prediction has been confirmed already, if by $n(k)$ we 
understand the slope of the initial spectrum $P_0(k)$ smoothed over the range 
$\Delta \ln k \sim 1$. To see this, it is even not neccessary to use
results for $n$ following from the COBE experiment (though they also tell
us the same), it is sufficient to compare the COBE normalization of
perturbations for scales of the order of the present cosmological horizon
$R_h$ with the $\sigma_8$ normalization that follows, e.g., from the present
cluster abundance. The difference in the amplitude of initial perturbations 
at these two scales which are divided by approximately $3$ orders of magnitude 
is only $2-2.5$ times for the pure CDM model and even less for
other models, e.g., the $\Lambda$CDM model. In addition, these numbers give 
us an idea about the magnitude of expected deviations from the exact
scale invariance: $|n-1|\le 0.3$ (once more, we are speaking about a 
smoothed $n$). Observational effects related to the part of $P_0(k)$ 
between these two points (CMB temperature fluctuations at 
medium and small angles, galaxy-galaxy and cluster-cluster correlations,
peculiar velocities of galaxies) also do not require larger smooth deviations 
from $n=1$. 

Still, it is well established now that the simplest cosmological model of the 
present Universe - the CDM model with the $n\approx 1$ initial spectrum of 
adiabatic perturbations (SCDM) - does not work. So, we have
to go further and to introduce new elements (= new physics) into basic
Lagrangians describing either inflation in the early Universe or the
present dark matter content in the Universe. In the former case we change
$P_0(k)$. In the latter case we get a different
dynamics of expansion of the Universe at recent time and change both a 
matter transfer function and a law of perturbation growth. Fortunately,
the amount of required additional new physics can be parametrized by a 
few (1-2) new fundamental constants (in this respect, see the classification 
of cosmological models of the present Universe in \cite{clas}). Let me 
further discuss two interesting concrete possibilities. These new optional
possibilities is what further development of inflationary ideas gives for the 
present-day cosmology and LSS - this explains the title of my talk.   

\section{How to produce steps and spikes in $P_0(k)$}

The observational fact that the smoothed slope $n$ cannot be significantly
different from 1 does not exclude the possibility of {\em local} strong
deviations from the flat spectrum, i.e., steps and/or spikes in $P_0(k)$. 
Of course, one should not expect such a behaviour to be typical,
we shall see below that if it happens at all, it occurs at some preferred 
scales which themselves become new fundamental parameters
of a cosmological model. Do we have any observational evidence for an
existence of such preferred scales at the Universe? At present, only one
scale in the Fourier space, $k=k_0=0.05h$ Mpc$^{-1}$, remains a candidate for 
this role, and it seems that the spectrum is smooth for larger $k$ (from
galaxy-galaxy correlation data) and for smaller $k$ (from CMB data). 
Here $h$ is the present Hubble constant in terms of $100$ km/s/Mpc. As for
this scale itself, there exists an evidence for a peculiar behaviour (in the 
form of a sharp peak) in the Fourier power spectrum of rich Abell - ACO 
clusters (with richness class $R\ge 0$ and redshifts $z\le 0.12$) around it 
\cite{ein}. This anomalous behaviour persists if the distant border for the
cluster sample is reduced to $z=0.07-0.08$ \cite{retz}. If we assume that the 
cluster power spectrum is proportional to the power spectrum of the whole
matter in the Universe (with some constant biasing factor), and calculate
the corresponding {\em rms} multipole values $C_l$ of angular fluctuations
of the CMB temperature, they appear to be in a good agreement with existing
results of medium-angle experiments \cite{atrio}. Moreover, if $\Omega_m=1$,
the peak in the power spectrum inferred from the cluster data just explains 
an excess in $C_l$ for $l=200-300$ observed in the Saskatoon experiment.

On the other hand, there is no peak at $k=k_0$ in the power spectrum of both
APM clusters (which are generally less rich than Abell-ACO clusters) and
APM galaxies \cite{tad} (though some less prominent feature at this scale
may still exist in the latter spectrum \cite{cast}), and the maximum in these
spectra seems to be shifted to $k\sim 0.03h$ Mpc$^{-1}$. Leaving a solution
of this discrepancy to more complete future surveys, let us consider 
theoretical predictions.
 
It is possible to produce local features in the initial spectrum even
remaining (at least, formally) inside the standard paradigm of one-field 
inflation. The only thing which should be relaxed is the requirement of the
analyticity of an inflaton effective potential $V(\vph)$ at all points. So, 
let me admit that $V(\vph)$ has some kind of discontinuity at a point 
$\vph=\vph_0$. Of course, really this discontinuity is smoothed in a very
small vicinity of $\vph_0$. Three cases are the most interesting.

1. $[V]=[V']=0,~[V'']\not= 0$ at $\vph=\vph_0$.

\noindent
Here $[]$ means the jump in the quantity considered, namely,
$[A]\equiv A(\vph_0 +0)-A(\vph_0-0)$, and the prime denotes the derivative
with respect to $\vph$. If we assume that the slow-roll
conditions $V'^{2}\ll 48\pi GV^2,~|V''|\ll 24\pi GV$ are satisfied near the 
point $\vph=\vph_0$ (here and below $c=\hbar=1$), then 
in the zero-order approximation the standard result for a perturbation 
spectrum is valid:
\be
P_0(k)={k^4R_h^4(t)\over 400}h^2(k),~~k^3h^2(k)= 18\left({H^6\over V'^2}
\right)_k, ~~H\equiv {\dot a\over a}\approx \sqrt{{8\pi GV\over 3}},
\label{stand}
\ee
where the index $k$ means that the quantity is evaluated at the moment of the 
first Hubble radius crossing ($k=aH$) at the inflationary stage. The result 
for $P_0(k)$, in contrast to the metric perturbation $h^2(k)$ defined in the 
ultra-synchronous gauge ($h$ is equal to $1/3$ of the trace of a spatial 
metric perturbation in this gauge, see, e.g., the original paper 
\cite{star82}), refers to the matter-dominated stage where $R_h(t)=2/H=3t$. 
Note also that there is no necessity in adding the multiplier ${\cal O}(1)$ 
here.

So, in this case $P_0(k)$ is continuous at $k=k_0$ but its slope $n(k)$
has a step-like behaviour there (similar to the case considered in 
\cite{ein1}). However, due to small corrections to Eq. (\ref{stand}) which are 
beyond the slow-roll approximation, it appears that $n$ cannot be obtained 
simply
by differentiating (\ref{stand}), and I expect that the sharp behaviour in
$n$ will be smoothed near $k_0$. This question is still under consideration.

2. $[V]=0,~[V']\not= 0$ at $\vph=\vph_0$.

\noindent
Now the second of the slow-roll conditions is violated, while we can choose
parameters of the jump in such a way that the first condition is still valid.
Naive application of Eq.(\ref{stand}) would give a step in $P_0(k)$. However,
the slow-roll approximation is clearly not applicable. The exact solution for 
a local part of the spectrum near the point $k_0$ was obtained in 
\cite{star92}. It reads:
$$k^3h^2(k)={18H_0^6\over V_-'^2}D^2(y),~H_0=\sqrt{{8\pi GV(\vph_0)
\over 3}},~V'_{\pm}=V'(\vph_0\pm 0)>0,~y={k\over k_0}, $$
\be
D^2(y)= 1-3\left({V_-'\over V_+'}-1\right){1\over y}\left(\left(1-
{1\over y^2}\right)\sin 2y+{2\over y}\cos 2y\right)  
\label{step}
\ee
$$+{9\over 2}\left({V_-'\over V_+'}-1\right)^2{1\over y^2}\left(1+
{1\over y^2}\right)\left(1+{1\over y^2}+\left(1-{1\over y^2}\right)\cos 2y-
{2\over y}\sin 2y\right). $$      
The function $D^2(y)$ has a step-like behaviour with superimposed oscillations.
Since $D(0)=V_-'/V_+',~D(\infty)=1$, the spectrum approaches the flat spectrum
if $|\ln(k/k_0)|\gg 1$. As compared with the flat spectrum, the spectrum
(\ref{step}) has more power at large scales (small $k$) if $V_-'>V_+'$,
and more power on small scales in the opposite case. The shape of this 
function is universal (in the sence that it does not depend on a way of 
smoothing the jump in $V'$, if it is made in a sufficiently small vicinity of 
$\vph_0$), it depends on the ratio $V_-'/V_+'$ only.

3. $[V] \not= 0$ at $\vph=\vph_0$.

\noindent
In this case, there is no universal spectrum, and the answer depends on a
concrete way of smoothing $V(\vph)$ at $\vph =\vph_0$. Some general results
are presented in \cite{star92}, typically $P_0(k)$ acquires a large bump,
however, a well may appear, too. Fortunately, there is no need in further
consideration of this more complicated case, since observational data do
not require such a strong non-analyticity. Looking at the results presented
in \cite{ein,retz} it is clear that they lie somewhere between the first two
cases.

Therefore, a general lesson from these considerations is that pecularities in
$V(\vph)$ can produce local features in $P_0(k)$ where the slope $n$ is
significantly different from unity. However, these features cannot be too 
sharp, in particular, both $P_0(k)$ and $n(k)$ are expected to be continuous 
functions of $k$.

So far, the treatment of peculiar points was purely mathematical. However,
if we are seeking for a physical explanation of such behaviour of $V(\vph)$,
we have to go beyond the paradigm of one-field inflation to a more
complicated case of two-field inflation. Inflation with two scalar field is
a very rich physical model which includes double inflation, hybrid inflation,
open inflation, etc. as specific cases. In our case it is sufficient to 
assume that the second scalar field $\chi$, in contrast to the inflaton
field $\vph$, is always in the regime $|m_{\chi}^2|\gg H^2$. So, it is not
dynamically important during the whole inflation. However, it is coupled to
$\vph$ (e.g., through the term $g^2\vph^2\chi^2$) and, as a result
of change in $\vph$ during inflation, the field $\chi$ experiences a fast
phase transition approximately $60$ e-folds before the end of inflation.
If parameters of an interaction potential $V(\vph,\chi)$ are such that the
phase transition may be considered as an equilibrium one, then its net effect
on inflation appears in the change of the equilibrium effective potential 
$V_{eff}(\vph)\equiv V(\vph,\chi_{eq}(\vph))$ only. If the transition is a 
second-order 
one, with no jump in $\chi_{eq}$, the first case considered above takes 
place. If the transition is a first-order one, with a non-zero jump in 
$\chi_{eq}$, we arrive to the second case.

\section{How to determine a variable cosmological term from observations}  

There exists a growing amount of evidence that the total energy density
of matter in the Universe including baryonic and nonbaryonic dark matter
is significantly less than unity ($\Omega_m \sim 0.3 - 0.4$). The most
recent argument for this conclusion is based on the evolution of
abundance of rich galaxy clusters with redshift $z$ \cite{bah}. Of course,
as usually in modern cosmology, observational papers with the opposite 
conclusion have appeared almost immediately \cite{blan} (once more, I leave
the solution of this dilemma for future).
A natural way to incorporate $\Omega_m <1$ without abandoning the paradigm
of one-field inflation is to assume the existence of 
a positive effective cosmological constant with the energy density (in terms 
of the critical one) $\Omega_{\Lambda}=1-\Omega_m$. This leads to the flat
$\Lambda$CDM cosmological model with the $n\approx 1$ initial spectrum of 
adiabatic perturbations which is in a good agreement with all existing 
observational data for a rather large region in the space of parameters 
$H_0,~\Omega_m$.

It is clear that the introduction of a cosmological constant requires
new and completely unknown physics in the region of ultra-low energies.
If we try to describe it phenomenologically by the same kind of physics
which was so successively used in the paradigm of one-field inflation, 
namely, by a scalar field with some interaction potential $V(\vph)$ minimally 
coupled to the Einstein gravity, then the conclusion is that generally 
this "constant" may be weakly time-dependent. Models with a time-dependent
cosmological constant were introduced ten years ago
\cite{wet}, and different potentials $V(\vph)$ (all inspired by inflationary
models) were considered: exponential \cite{wet1}, inverse power-law
\cite{rat}, power-law \cite{weiss}, cosine \cite{friem}.

However, it is clear that since we know essentially nothing about physics
at such energies, there is no preferred theoretical candidate
for $V(\vph)$. On the other hand, using the cluster abundance $n(z)$
determined from observations and assuming the Gaussian statistics of
initial perturbations (the latter follows from the paradigm of one-field 
inflation, too, and is in agreement with other observational data), it is 
possible to determine $\delta (z)$ - the time evolution of the linear 
density contrast in the CDM component for a fixed comoving scale 
$R\sim 8(1+z)^{-1}h^{-1}$ Mpc up to $z\sim 1$, using the Press-Schechter
approximation. It can be shown that the knowledge of $\delta (z)$ 
uniquely determines the form of the effective potential $V(\vph)$ required 
for this model (simultaneously, the present value of $\Omega_m$ is determined
from the same data, too). Therefore, I propose not to introduce $V(\vph)$ 
by hand, but to find it from data on clustering of high-redshift objects. 
Details will be published separately. Then a completely independent test 
of the model is provided by CMB temperature anisotropies.

\end{document}